\documentclass[aps,reprint]{revtex4-1}

\pdfoutput=1

\usepackage[centertags]{amsmath}
\usepackage{amsfonts}
\usepackage{amssymb}
\usepackage{amsthm}
\usepackage{amscd}
\usepackage[pdftex]{graphicx}

\newcommand{\idn}{{1\relax{\kern-.35em}1}}

\newcommand{\qmark}[1]{\accentset{\relax{\kern+.5em}?}{#1}}

\newcommand{\sla}{\relax{\kern-.5em{\backslash}}}

\newcommand{\ads}[1]{AdS$_#1$}

\newcommand{\pd}{\partial}

\newcommand{\s}{\sigma}
\renewcommand{\d}{\delta} 
\newcommand{\eps}{\varepsilon}

\newcommand{\tr}{\textup{Tr}~}

\newcommand{\abs}[1]{\left\vert~#1~\right\vert}

\newcommand{\ave}[1]{\left\langle #1 \right\rangle}

\newcommand{\zbra}[1]{\Bigl\langle\relax{\kern-.4em}\Bigl\langle~#1~\Bigr\vert}
\newcommand{\zinnp}[2]{\Bigl\langle\relax{\kern-.4em}\Bigl\langle~ #1~\Bigm\vert~ #2~\Bigr\rangle}            
\newcommand{\zbok}[3]{\Bigl\langle\relax{\kern-.4em}\Bigl\langle ~#1~\Bigl\vert~ #2~\Bigr\vert~#3~\Bigr\rangle} 

\renewcommand{\star}{*}

\begin{document}
\title{A Holographic Chiral $p_x+ip_y$ Superconductor}
\author{Leopoldo A. Pando Zayas}
\email{lpandoz@umich.edu}
\author{Dori Reichmann}
\email{dorir@umich.edu}
\affiliation{Michigan Center for Theoretical Physics, University of Michigan, Ann Arbor, MI 48109}
\begin{abstract}
Using a string inspired model we construct gravity duals generalizing $p_x$ and $p_x+ip_y$ superconductors. Introducing a Chern-Simons coupling in the gravity side we demonstrate the ability to control which phase dominates at low temperatures, and focus on the chiral $p_x+ip_y$ phase. We study the fermionic spectral function and establish that the behavior is rather different from the standard p-wave two-nodes model.
\end{abstract}
\preprint{MCTP-11-31}

\maketitle

{\it\bf Introduction -- }
The discovery of materials with possible chiral p-wave superconductivity has gathered much interest in condensed matter physics. This type of unconventional superconductors \cite{KITPTALK,Mackenzie2000148, PhysRevLett.104.147001} features a complex vector order parameter (abbreviated as the $p+ip$ wave), time reversal breaking, and a complex gap function with the phase acquiring a $2\pi$ shift when rotated around the symmetry axis, {\it i.e.}, $\Delta_\pm(\mathbf{k})\propto (k_x\pm ik_y)$. In theoretical models $p+ip$ superconductivity is also related to electron-hole symmetry and the  breaking of spin rotation. Chiral p-wave superconductivity was reported in Sr$_2$RuO$_4$  \cite{Nelson12112004, RevModPhys.75.657}, some heavy fermion compounds \cite{Strand11062010, RevModPhys.56.755}, and was suggested in the study of of the Iron-pnictide superconductors (SC) \cite{PhysRevLett.101.057003, PhysRevB.78.144517, 2009JPSJ...78f2001I}.

The AdS/CFT correspondence \cite{Maldacena:1997re, Gubser:1998bc, Witten:1998qj} provides calculational tools for approximations in strongly coupled field theories, inspiring a search for connections between string theory and low energy physics of compound materials like the cuprates, graphene,  non-Fermi liquids and more.

{\it\bf Review -- }There are three main models to study unconventional superconductivity via the AdS/CFT duality:
The Abelian Higgs model \cite{Hartnoll:2008kx, Hartnoll:2008vx, Gauntlett:2009dn}, the non-Abelian model \cite{Gubser:2008wv,Gubser:2008zu}, and the charged spin 2 model \cite{Chen:2010mk, Benini:2010qc, Chen:2011ny}. The distinctive feature of these models is the order parameter symmetries: s-wave, p-wave and d-wave respectively. Each of  the models has its advantages and disadvantages both on the pure theoretical side and when applied to real condensed matter systems. These models and their deformations span the dictionary between string theory and superconductors in condensed matter physics.	

The p-wave model exhibits a vector order parameter and two superconductor phases, with order parameters $\ave{p_x}$ and $\ave{p_x+ip_y}$. In the original model, and all previous generalizations, the non-chiral $\ave{p_x}$ is the only thermodynamically stable phase. In this letter we focus on a natural, from the string theory point of view, extension of the non-Abelian holographic superconductor which exhibits $p+ip$ order parameter as the thermodynamically dominant phase. The extension is based on the introduction of an anomalous global symmetry in the field theory  which corresponds to a Chern-Simons term on the gravity side.  This particular modification descends directly from considering gauged supergravities and has already shown its potential as a description of a quantum critical point \cite{D'Hoker:2009bc}.

{\it\bf The $\ave{p_x}$ and $\ave{p_x+ip_y}$ Holographic Superconductors --} The p-wave holographic superconductors studied in \cite{Gubser:2008wv,Gubser:2008zu,Ammon:2009xh,Basu:2009vv,Ammon:2010pg,Aprile:2010ge} are framed in a phenomenological model that contains a dimensionless tuneable parameter $\alpha$ measuring the relative strength between the AdS radius and the gauge group coupling. The unbroken phase is the Reissner-Nordstrom-Anti de Sitter (RNAdS) black hole. At zero temperature this is the extremal solution. To identify the potential instabilities it is worth discussing the $AdS_2$ geometry arising in the near horizon region of the extremal solution. The competition controls the effective mass of a non-Abelian mode in the near horizon geometry. A strong enough YM coupling can generate a negative effective mass below the Breitenlohner-Freedman (BF) bound of $AdS{}_2$  and thus drive an instability and a global phase transition.

Gubser's model in D-dimensions can be summarized by the Einstein-Yang Mills action with $SU(2)$ gauge group,
\begin{align}\label{action0}
    S_0 =& \frac1{4\kappa_D^2}\int\bigg[\left(R+2\Lambda_D\right)\star\idn-\frac1{2}F^i_{(2)}\wedge\star F^i_{(2)}\bigg],
    \\
    F_{(2)}^i =& dA^i_{(1)}+\frac{\alpha}{\sqrt 2 L}\,\eps^{ijk}A^j_{(1)}\wedge A^k_{(1)},
\end{align}
with $i,j=1,2,3$ and the cosmological constant $\Lambda_D = -\frac{(D-1)(D-2)}{2L^2}$. There are two possible types of non-Abelian condensate distinguished by the value of the 4-form $F^i_{(2)}\wedge F^i_{(2)}$. For the $\ave{p_x}$ case the 4-form vanishes, while for the $\ave{p_x+ip_y}$ it has a finite value. The gauge field equation of motion prohibits turning on these two modes simultaneously.

%
%
We modify Gubser's model in 5 dimensions by introducing a Maxwell field and a Chern-Simons term:
\begin{align}\label{action}
    S =& S_0+S_1,\\
    S_1=&-\frac1{8\kappa_5^2}
    \int\bigg[G_{(2)}\wedge\star  G_{(2)}+\gamma F^i_{(2)}\wedge F^i_{(2)}\wedge B_{(1)}\bigg],
    \\
    G_{(2)}=&dB_{(1)}.
\end{align}
The form of the modification is inspired by 5-dimension $SU(2)\times U(1)$ gauged supergravity (which is a consistent reduction of type IIB string theory). However, the full supergravity contains additional neutral scalars, and our constants $\gamma$ and $\alpha$ are replaced by scalar functions.

On the gravity side, the Chern-Simons term induces an explicit breaking of the time-reversal symmetry.
In the holographic field theory dual the Chern-Simons term introduces a 'tHooft anomaly between the $SU(2)$ and $U(1)$ global currents. The parameter $\gamma$ measures the strength of the anomaly and it is fixed by the dynamical properties of the field theory. In order to interpret these solutions as superconductors, one needs to weakly gauge the Cartan of the $SU(2)$, i.e. $A^3_{(1)}$ (this is the field theory photon). The anomaly will then break the $U(1)$ global symmetry, and it cannot be identified as a conserved current in the IR field theory. The additional two non-Abelian currents ($A^1_{(1)}$ and $A^2_{(1)}$) remain as conserved currents in the IR. The appearance of the $SU(2)$ currents in the field theory forces electron-hole symmetry for any charged fermionic excitation. Alternatively, the currents can be interpreted as enhancement of electron-hole symmetry in the low-energy effective action.

Note that the $\ave{p_x}$ solution is simply embedded in the new model in a $\gamma$ independent manner. Since $F^i_{(2)}\wedge F^i_{(2)}=0$ there will be no source to Maxwell field $B_{(1)}$ and without external sources it will vanish $G_{(2)}=0$.

We can gain intuition into the effect of the Chern-Simons term by studying the model near an \ads2$\times \mathrm{R}^3$ throat (the near horizon of an extremal RN solution). The solution Ansatz is
\begin{align}\label{Ansatz-ads2}
    ds^2_{AdS_2} =& -\frac{\rho^2}{l^2}dt^2+\frac{l^2}{\rho^2}d\rho^2+\frac{r_h^2}{12l^2}d\vec x^2,
    \\
    A^i_{(1)}=&-\d^{i3} \frac{\sqrt2\rho}{l}dt + \d^{i1} \zeta(\rho)dx+\cr&+\omega(\rho)\left(\d^{i2}dx+\d^{i1}dy\right),
    \\
    B_{(1)}=&\,p(\rho)dz.
\end{align}
Where $\zeta(\rho)$ and $\omega(\rho)$ are respectively the $\ave{p_x}$ and $\ave{p_x+ip_y}$ modes.
The linearized equations of motion for $p(\rho)$, $\omega(\rho)$ and $\zeta(\rho)$
\begin{align}
    p''(\rho) =&\,0,
    \cr
    \left[\rho^2\zeta'(\rho)\right]' +\frac{\alpha^2}{3}\zeta(\rho)=&\,0,
    \cr
    \left[\rho^2\omega'(r)\right]'+\frac{\alpha^2}{3}\omega(\rho)=&\,0.
\end{align}
Notice that the only parameter governing the equation is the dimensionless parameter $\alpha$. We find an instability for both non-Abelian modes (comparing the mass to the BF bound of \ads2) if
\begin{equation}\label{critical-alpha}
    \alpha \geq \alpha_c = \sqrt{3}/2.
\end{equation}
Remaining in the probe limit, adding the non-linear terms we find
\begin{equation}
\zeta(r) p'(r) =0, \quad \zeta^2(r)\omega^2(r) =0,
\end{equation}
which is the statement that the two condensates cannon coexist. Focusing on the $\ave{p_x+ip_y}$ condensate, still in the probe limit, the general equations become
\begin{align}
    & \left[p'(\rho)-\frac{\gamma\alpha}{6l\rho}\omega^2(\rho)\right]' =0
\\
    &\left[\rho^2\omega'(r)\right]'
    +\left[\frac{\alpha^2}{3}-\frac{\alpha^2}{72r_h^2l^2}\omega^2(\rho)-\frac{\gamma\alpha\rho}{6r_h}p'(\rho)
    \right]\omega(\rho)=0
\end{align}
For non-zero $\gamma$ an instability of $\omega(\rho)$ will source the Maxwell field which in turn backreacts on $\omega(\rho)$. There is no reliable probe limit for $T=0$, so to find an honest solution we need to take into consideration the equation for the metric and the charge density $A^3_{(1)}$.

The most urgent question is whether the Chern-Simons term can change the thermodynamics of the system. There is no analytical answer for this question and we are forced to turn our focus to numerical solutions. When discussing thermodynamic stability it is important to consider the full solution (not only the probe limit), as back-reaction can change the nature of the phase transition \cite{Ammon:2009xh} .

{\it Numerical solutions --} We solve the equations of motion numerically (beyond the probe limit), and scan the effect of $\gamma$, for simplicity we set the \ads5 scale $L=1$. We use the following Ansatz
\begin{align}\label{Ansatzb}
    ds^2 =& -r^2f(r)H^2(r)dt^2+\frac{dr^2}{r^2f(r)}
    +\cr&
    +r^2G^2(r)(dx^2+dy^2)
    +\frac{r^2}{G^4(r)}dz^2,
    \cr
    A^i_{(1)} =& u(r)\d^{i3}dt+\zeta(r)\d^{i1}dx+\omega(r)\left(\d^{i2}dx+\d^{i1}dy\right),
    \cr
    B_{(1)} =& p(r)dz.
\end{align}
The leading terms of the boundary expansions are:
\begin{align}
&    r\rightarrow\infty~:
&&
    G(r) \rightarrow 1+\frac{G_1}{r^4},
&&
    f(r) \rightarrow1+\frac{M_1}{r^4},
\cr
&&&
    H(r) \rightarrow 1,
&&
    u(r) \rightarrow U_0+ \frac{U_1}{r^2},
&
\cr
&&&
    \zeta(r) \rightarrow \frac{Z_1}{r^2},
&&
    \omega(r) \rightarrow \frac{W_1}{r^2},
&
\cr
&&&
    p(r) \rightarrow \frac{P_1}{r^2},
&&&
\end{align}
$U_0$ and $U_1$ are (proportional) to chemical potential and charge density of the solution. We used the scaling symmetries to set the asymptotic value of $G$, $H$ and $f$ to 1. We also set the non-normalizable modes of $\zeta$, $\omega$ and $p$ to zero leaving the corresponding normalizable modes $Z_1$, $W_1$ and $P_1$ free. Finally there are two additional normalizable modes $M_1$ and $G_1$ related to metric terms.

The leading terms of the near horizon expansions are:
\begin{align}
&    r\rightarrow r_h~:
&&
    H(r)\rightarrow S_t,
&&
    G(r)\rightarrow \frac1{\sqrt{S_{x}}},
    &
\cr
&&&
    f(r)\rightarrow f_0\frac{(r-r_h)}{r_h},
&&
    u(r)\rightarrow S_tu_0(r-r_h),
    &
\cr
&&&
    \omega(r)\rightarrow S_{x}\omega_0,
&&
    \zeta(r)\rightarrow S_{x}\zeta_0,
&
\cr
&&&
    p(r)\rightarrow\frac{p_0}{S_{x}^2},
&&&
\end{align}
with $f_0=4-\frac{u_0^2}{12\alpha^2}-\frac{\omega_0^4}{12\alpha^2r_h^4}$.
The parameters $S_x$ and $S_t$ are related to scaling symmetries, the scaling is fixed at the boundary expansion so we need to follow it in the near horizon expansion (and cannot fix it). Regularity in near horizon eliminates potential $\log(r-r_h)$ terms in the expansion. The elimination of a constant term for $u(r)$ fixes the gauge choice, to the gauge where $U_0$ is interpreted as the chemical potential. The vector hair at the horizon is given by $\zeta_0$, $\omega_0$ and $p_0$. The three types of solutions we consider are: (1) $\omega_0=p_0=0$ (2) $\zeta_0=0$ and (3) Abelian solution with all hair vanishing. These solutions correspond respectively to $\ave{p_x}$, $\ave{p_x+ip_y}$ and no condensate phases.

We solve the equations by using a two section shooting algorithm, matching the near horizon and the \ads5 boundary expansions. We use numerical solutions in the intermediate sections to integrate the equation for a near horizon initial radius $r_i$ and asymptotic initial radius $r_f$ to an intermediate value $r_m$. At the intermediate point we define a multi-dimensional goal function as the difference between all variables (and first derivatives). The goal function is a 12-dimensional function of the 14 initial condition parameters at $r_i$ and $r_f$. Using a Newton-Raphson algorithm we find a root of the goal function fixing all parameters as functions of the black hole chemical potential, the black hole temperature and the Lagrangian parameters $\alpha$, and $\gamma$. We further set the chemical potential such that $U_0=1$, thus working in a grand-canonical ensemble and measuring all quantities in units of the chemical potential.

\begin{figure}[ht]
\begin{center}
\includegraphics[width=80mm]{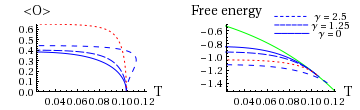}
\end{center}
\caption{\label{fig:thermal}Dependence of the condensate and free energy on temperature (arbitrary units). Continuous green line is the regular phase, dotted red line is the $\ave{p_x}$ phase, and dashed blue lines are $\ave{p_x+ip_y}$. Plots are for $\alpha=1.6$ and the dashed curved realetd to $\gamma=0$, $\gamma=1.25$ and $\gamma=2.5$.}
\end{figure}

{\it Regularization of the Euclidean action - }The regularized Euclidean action is \cite{Emparan:1999pm}
\begin{align}
    I_{reg} = I_{bulk}+I_{surf}+I_{ct},
\end{align}
where the Euclidean bulk action is the analytic continuation of \eqref{action}.
The surface term is the Gibbons-Hawking boundary term. For a Yang-Mills Einstein Gravity the only relevant counter term is related to the induced metric on the boundary since the solution of interest has no divergence coming from the Yang-Mills part.
\begin{equation}
    I_{surf}+I_{ct} = \frac1{4\kappa_5^2}\int_{\pd M}d^4x\sqrt{h} \left(\frac{3}{L}-h^{ij}\nabla_i n_j\right)
\end{equation}
with $h$ the induced metric on the boundary and $n$ is an outward pointing unit normal vector to $\pd M$. Cutting the geometry at a large radius $r=r_+$, we find
\begin{multline}
    I_{surf}+I_{ct} =\frac1{4\kappa_5^2}\int_{\pd M}d^4x~\frac{r_+^4}{L^2}H(r_+)\\\sqrt{f(r_+)}\left[3-2\,\pd_r\left(r\sqrt{f(r)}\,\right)\right]_{r=r_+}
\end{multline}
{\it Results - } We find numerically that $G_1=P_1=0$, so the non-vanishing normalizable modes are related to the energy density($M1$), charge density($U_1$) and the two possible condensates $Z_1$ and $U_1$. Figure \ref{fig:thermal} presents a typical thermal behavior of the solutions. We find that increasing the value of $\gamma$ lowers the free energy of the $\ave{p_x+ip_y}$ solutions until the point where they win the thermodynamic battle against the $\ave{p_x}$ solutions. We also see that large values of $\gamma$ (around 2.5) can change the nature of the phase transition from second order to first order. We note that at smaller values of $\alpha$ (than those shown in plots) the $\ave{p_x}$ solution also exhibits first order phase transition, while the $\ave{p_x+ip_y}$ solution is always second order at $\gamma=0$.

\begin{figure}[ht]
\begin{center}
\includegraphics[width=80mm]{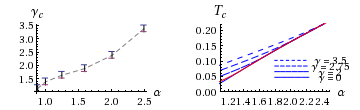}
\end{center}
\caption{\label{fig:gammac}On the right side we plot the dependence of the critical temperature $T_c$ on the gauge coupling strength $\alpha$. The continuous red line is the $\ave{p_x}$ phase, and the dashed blue lines are $\ave{p_x+ip_y}$ with $\gamma=0,2,2.75,3.5$. On the left we plot the dependence of $\gamma_c$ on $\alpha$. The critical value $\gamma_c$ is defined as the minimal value of $\gamma$ where the $\ave{p_x+ip_y}$ dominates at $T=0.5T_c$ (we use the critical temperature of $\ave{p_x}$).}
\end{figure}

The summary of the dependence of the phase diagram on $\gamma$ and $\alpha$ is given in \ref{fig:gammac}. On the right side we see the dependence of the critical temperature on $\alpha$. The left side holds the answer for the most urgent question, we see that for any value of $\alpha$ (in the range tested) there exist a critical value of $\gamma$ above which the $\ave{p_x+ip_y}$ is the thermodynamically favored phase. We define $\gamma_c$ by comparing the free energy at $T=0.5T_c$, using $T_c$ of the $\ave{p_x}$ solution. The error bars on the plot are the result of the heavy computation time involved in finding each value, and not machine precision.

The near horizon Ansatz we use is valid for finite temperature solutions only, leading to a minimal temperature below which we loose numerical accuracy. Therefore the question of which phase dominates at $T=0$ remains open. However for large enough values of $\gamma$ it is likely that the $\ave{p_x+ip_y}$ phase will remain thermodynamically favored all the way down to zero temperature.

{\it\bf The Spectral Function --} Angle Resolve Photoemission Spectroscopy (ARPES) allows to directly probe the fermionic structure of materials. In our theoretical model the fermion spectral function can be read from the retarded Green function of probe fermions
\begin{equation}
    \rho(\omega,\vec k) = \frac1{\pi}\mathrm{Im}\left[\tr G_R(\omega,\vec k)\right]
\end{equation}
where $\omega, \vec k$ are frequency and momentum in the dual field theory.

We use a ``minimally coupled'' fermion with zero mass transforming in the $\mathbf{\underline 2}$ of the $\mathrm{SU}(2)$ gauge symmetry and with charge $q=1/2$ under the $U(1)$ gauge symmetry. See appendix-\ref{appx-ferm} for details. Rotations around the axis of symmetry ($z$) are canceled by $SU(2)$ rotations, and after the trace the spectral function is invariant in the $x-y$ plane. We use polar coordinates in the $x-z$ plane $k_z = k\cos\theta$. We calculated numerically the spectral function for $\alpha=1.59$, $\gamma = 2.5$ and temperature $T/T_c = 0.2$. The results are displayed in figures \ref{fig:spcfunc}-\ref{fig:edc}. For comparison in appendix-\ref{appx-px} we reproduce the same results for the $\ave{p_x}$ SCs.

In figure-\ref{fig:spcfunc} we can see the general form of the spectral function in the $k-\omega$ plane at a fixed angle ($k_x/k_z =\tan\theta$). The spectral function is invariant under reflection in the frequency $\omega\rightarrow-\omega$. Note that this symmetry is merely a consequence of the trace but in general it will be broken.  We can clearly see two separate excitations (resonances), each appearing twice (negative and positive frequency). The Fermi-momentum ($k_F$) for each excitation is the momentum where the excitation frequency is closest to zero. Half the frequency distance between the positive and negative excitation (of the same type) at the Fermi-momenta is the excitation gap \footnote{Not to be confused with the global gap, which is the range of frequencies where there are no states at all (vanishing spectral function)}.

\begin{figure}[ht]
\begin{center}
\includegraphics[width=80mm]{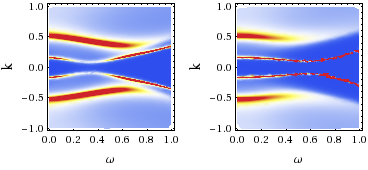}
\end{center}
\caption{\label{fig:spcfunc} A density map of the spectral function at fixed direction in the $x-z$ plane. The color scheme changes from near zero values in blue to high values in red. The plot is clipped below the maximal value of the spectral function to allow to see the details better. On the left we look at $\theta =\frac{5\pi}7$ and $\theta =\frac{\pi}{2}$ on the right. }
\end{figure}

Figure-\ref{fig:gapfunc} follows the change of the excitation gap and Fermi momentum for both excitations as we change the angle $\theta$. The higher frequency excitation is gapped at all angles, and the gap size changes mildly with angle, while the corresponding Fermi momentum ($k_F^{high}$) moves from zero near $\theta=0$ to $~0.6$ at $\theta=\pi$. The lower frequency excitation is not gapped at $\theta=0$ and develops a gap as the angle increases, the corresponding Fermi momentum ($k_F^{low}$) changes mildly with $\theta$. As can be seen from left plot in figure-\ref{fig:spcfunc} the lower frequency excitation, can be strongly attenuated at $k_F^{low}$. We do not have a clear explanation for this behavior, and it may be the result of the interplay between the two resonances.

\begin{figure}[ht]
\begin{center}
\includegraphics[width=80mm]{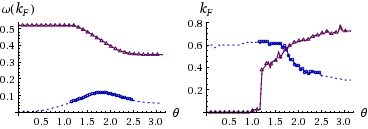}
\end{center}
\caption{\label{fig:gapfunc} The excitation gap (left) and Fermi momentum (right) for the two excitations as a function of angle $\theta$. The blue circle are related to the lower frequency excitation and the red triangles for the high frequency excitation. The lower frequency excitation (blue circles), shows a strong attenuation near it's $k_F$ for a wide range of angles. When the attenuation is below $10^{-3}$ (arbitrary cutoff) we replace the circles by dots.}
\end{figure}

To further study the angle dependance of the spectral function we study the Energy Distribution Curve (EDC), which is defined as the spectral function at the Fermi-momentum. In Figure-\ref{fig:edc} we plot the EDCs for both Fermi-momenta. The plot on the left, which is the EDC at $k_F^{low}$, shows that the peak is strongly attenuated for angles near the poles ($\theta = 0$ and $\pi$). The EDC at $k_F^{high}$ show the high frequency peak disappears at mid range angles, however this is not due to attenuation near the Fermi-momenta, but due to the dominace of the lower frequency peak at these range of $\theta$.

\begin{figure}[ht]
\begin{center}
\includegraphics[width=80mm]{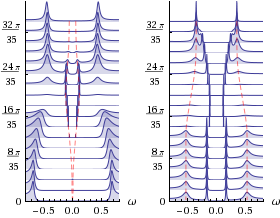}
\end{center}
\caption{\label{fig:edc} The EDC (spectra function at the Fermi-momenta) as function of frequency. The different plots are related to different angles $\theta$. The left and right plots are the EDC at the $k_F^{low}$ and $k_F^{high}$ correspondingly. The dashed red line represent the location of the gap. The different EDC's are normalized differently between the angles (for this plot only) and should not be directly compared. }
\end{figure}

In appendix-\ref{appx-xtra} we elaborate upon additional properties of the spectral function. First on the isospin current and then on the node structure. The isospin current is just the consequence of the relation between spatial rotation and rotation of the $\mathrm{SU}(2)$ group. Rotation of the SC solution in the x-y plane is not an exact symmetry, but can be compensated by a global rotation in the gauge group $\mathrm{SU}(2)$. This property can be detected in the fermionic Green's function by considering
\begin{equation}
    I^{(isospin)}(\vec k) = -\frac{i}{\pi}\sum_\omega\tr\left[G_R\left(\omega,\vec k\right)\tau^1\right].
\end{equation}
The node structure is studied by examining the low frequency spectral function, for the $\ave{p_x+ip_y}$ we see a single node near $\vec k=(0,0,-0.57)$,

{\it\bf Discussion --}
To summarize, in the first part we showed how a holographic chiral $p+ip$ superconductor can be constructed using the full treatment of gravity equations of motion.  In the second part we study the fermionic spectral function which diplay some unusual behavior compared to the $p$-wave superconductors (see appendix-\ref{appx-px}), the new features found:
\begin{itemize}
\item Two competing modes {\it versus} a single dominant mode in the $\ave{p_x}$ SCs
\item Chirality, i.e relation between spatial and $\mathrm{SU}(2)$ rotations which is absent in the $\ave{p_x}$ SCs.
\item Single node spectral function.
\end{itemize}

It would be interesting to pursue some of the following directions. For example, it would be desirable for condensed matter applications to consider similar constructions where the time-reversal symmetry is broken spontaneously. However, this would likely require a non-radial solution on the gravity side.

Recently, Son and Surowka \cite{Son:2009tf} have analyzed the hydrodynamics in the presence of an anomalous current and its description in terms of gravity. It would be interesting to use their approach to analyze our gravity solutions which should lead to the hydrodynamics of a $p+ip$ superconductor.

{\it \large Acknowledgments -- }
 We thank R. Leigh and J. Sonner for various explanations and comments. This work is  partially supported by Department of Energy under grant DE-FG02-95ER40899 to the University of Michigan.

\appendix
\section{Calculation of the spectral function}\label{appx-ferm}

The tree level contribution for the Green function is calculated by solving the Dirac equation of a probe fermion in the black hole background with in-falling boundary condition at the horizon. We use a ``minimally coupled'' fermion with mass $M$ transforming in the $\mathbf{\underline 2}$ of the $\mathrm{SU}(2)$ gauge symmetry and with charge $q$ under the $U(1)$ gauge symmetry.
\begin{align}
    \Gamma^a E_a^{\mu}&\left(\pd_\mu+\frac14\omega_{\mu ab}\Gamma^{ab}
    -iqB_\mu\right)\Psi_I+\cr&
    -M\Psi_I-i\frac{\alpha}{\sqrt2 L}\,\Gamma^a E_a^{\mu}\,A^i_{\mu}\tau^i_{IJ}\Psi_J
    =0,
\end{align}
where $\Gamma^a$ are flat space Dirac matrices, $E_a^\mu$'s are the inverse Vielbein, and $\tau^i$'s are generators of the $su(2)$ algebra.
We use the following redefinition of the fermion field $\Psi_I=(-g g^{rr})^{-\frac14}e^{-i\omega_0 t + ik_j x^j}\psi_I$. Near the \ads5 boundary the Dirac equation can be solved in terms of 4 normalizable  and 4 non-normalizable modes
\begin{align}
    \psi_I =& \sum_{\alpha=1}^2v^\alpha\,\left(r^{M/g}A_{I\alpha}+r^{-M/g-1}B_{I\alpha}\right)+\cr&
    +\sum_{\dot\alpha=1}^2u^{\dot\alpha}\,\left(r^{M/g-1}C_{I\dot\alpha}+r^{-M/g}D_{I\dot\alpha}\right)
\end{align}
where $A_{I\alpha},D_{I\dot\alpha}$ are free boundary conditions and $B_{I\alpha}$, $C_{I\dot\alpha}$ are constants fixed from the equation of motion. Note that $v^\alpha$ and $u^{\dot\alpha}$ are eigenvectors of $\Gamma^{\underline r}$ with eigenvalue $+1$ and $-1$ respectively.

In the near horizon limit (finite temperature) the metric takes the form
\begin{equation}
    ds^2 = -f_t^2(r-r_h)dt^2+\frac{f_r^2 dr^2}{(r-r_h)}+d\vec x^2.
\end{equation}
The Dirac equation solution behaves like
\begin{equation}
    \psi_I = \sum_{m=1}^4\left[a_m V^m_I(r-r_h)^{-i\frac{f_r\omega_0}{f_t}}
    +
    b_m U^m_I(r-r_h)^{i\frac{f_r\omega_0}{f_t}}\right],
\end{equation}
where $V^m_I$ and $U^m_I$ are a vector base for in-falling and out-going solutions respectively and $a_m$ and $b_m$ are free parameters. Since the equations are first order, setting in-falling boundary condition ($b_m=0$) impose a relation between the asymptotic boundary conditions. The asymptotic values are then determined (with a slight abuse of notations) as
\begin{equation}
    A = \underline{\textup{A}}a
    ~,\quad
    D = \underline{\textup{D}}a
\end{equation}
The retarded Green function is the $4\times 4$ matrix.
\begin{equation}
    G_R = -i\textup{\underline{D}}\,\textup{\underline{A}}^{-1}\Gamma^{t}
\end{equation}
Notice that $\Gamma^t$ should be projected onto the spin states, and it should not multiply the $SU(2)$ part of the Green function.

\section{Spectral function of the $\ave{p_x}$ superconductors}\label{appx-px}
In this appendix we reproduce the result for the spectral function of the $\ave{p_x}$ SCs in order to allow for a clear comparison between the two types of p-wave superconductors . To the best of our knowledge previous work \cite{Gubser:2010dm} covered only the probe limit, and the presentation of the result is different and harder to compare to the result of this paper.

\begin{figure}[ht]
\begin{center}
\includegraphics[width=80mm]{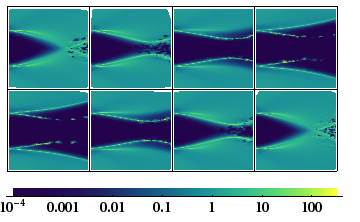}
\end{center}
\caption{\label{fig:app-spcZ} The spectral function of a $\ave{p_x}$ superconductor as a function of momentum (horizontal axis) and frequency (vertical axis). Plots corresponds to different angles $\theta =\tan^{-1}\frac{p_x}{p_y}$. The angle used on the top line from left to right are are $\theta=0,\frac{\pi}{7},\frac{2\pi}{7}$ and $\frac{3\pi}{7}$. The angles used on the bottom line from left to right are $\theta=\frac{4\pi}{7},\frac{5\pi}{7},\frac{6\pi}{7}$ and $\pi$. Unlike figure-\ref{fig:spcfunc} this plot use a logarithmic scale for the colors (see color table below the plot), which help to better compare the height of different peaks. The momentum range used is from $0$ to $1$ the frequency range used is from $-1$ to $1$ .}
\end{figure}

\begin{figure}[ht]
\begin{center}
\includegraphics[width=80mm]{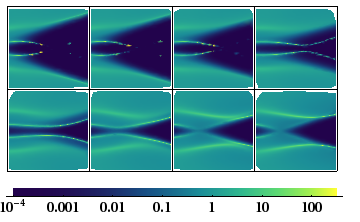}
\end{center}
\caption{\label{fig:app-spcW} The spectral function of a $\ave{p_x+ip_y}$ superconductor. See the caption of figure \ref{fig:app-spcZ} for details.}
\end{figure}

In figure \ref{fig:app-spcZ} we display the spectral function for a $\ave{p_x}$ superconductor at various directions in the momentum plane. Figure \ref{fig:app-spcZ} shows four quasi-particle excitations; two with positive frequencies and two with negative frequencies. Following the peak in the spectral functions we can find their dispersion relations $\omega(k)$. We can estimate their individual contributions to the whole spectral function by integrating the spectral function along these curves (spectral weight of a single excitation). There are also nodes near $\theta=0,\pi$ where there is no gaps in the spectral function $\int_k \rho(0,k)>0$. One can see the the inner excitation (closer to $k=0$) is dominating for all the angles, in the sense that it has higher spectral weight. Also one can see clear gapless excitation near the pole $\theta=0,\pi$. The spectral function show symmetry under rotation in the $y-z$ plane, and flipping of the frequency ($\omega\rightarrow-\omega$) and momentum along the symmetry axis ($k_x\rightarrow-k_x$). For comparison we plot the matchings plots of the $\ave{p_x+ip_y}$ superconductor in figure \ref{fig:app-spcW}. The qualitative difference is very clear, most prominently is the existence of two clear nodes near $\theta=0$ and $\pi$. A second clear difference is invariance of the $\ave{p_x}$ spectral function under $\theta->\pi-\theta$ (flipping the sign of the momentum). A more subtle difference is the dominance of the lower frequency excitation in the $\ave{p_x}$ case, in contrast to the $\ave{p_x+ip_y}$ spectral function where the two excitation alternate dominance with the angle and momenta (most obvious is the nearly disappearance of the lower frequency near $\theta=\pi$).

\section{Additional aspects of the Green's function}\label{appx-xtra}

In this section we give extra (numerical) details on the nature of the fermionic excitation as can be seen from the 2-point function.

\begin{figure}[h]
\begin{center}
\includegraphics[width=80mm]{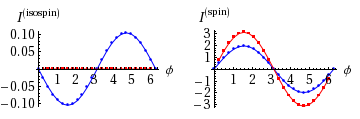}
\end{center}
\caption{\label{fig:app-current} The isospin (left) and spin (right) currents. The blue circles mark the $\ave{p_x+ip_y}$ superconductor, while the red squared mark the $\ave{p_x}$ superconductor. The angle $\phi$ is the angle around the symmetry axes $\tan^{-1}\frac{k_x}{k_y}$ and $\tan^{-1}\frac{k_z}{k_y}$ for the $\ave{p_x+ip_y}$ and $\ave{p_x}$ superconductors respectively. The plot are taken at momentum $\abs k = 0.88$.}
\end{figure}

The fermionic 2-point function is a $4\times4$ matrix, where the $4$ is combination of the global $\mathrm{SU}(2)$ symmetry and ($I=1,2$) and the projection over $\Gamma^r$ eigenvalues ($\alpha=1,2$). So far we studied the spectra function which is the imaginary par of the trace:
\begin{equation}\label{dos}
    \rho(\omega,\vec k) = \frac{1}{\pi}\mathrm{Im}\left[\tr G_R(\omega,\vec k)\right]
    =\frac1{\pi}\mathrm{Im}\sum_{I,\alpha}\left[G_R(\omega,\vec k))\right]_{I\alpha,I\alpha}
\end{equation}
The large $\mathrm{SU}(2)$ and spatial rotations transformations are non-gauged and appear as global symmetries in the field theory. The conserved currents corresponding to these symmetries is measurable in Josephson junctions experiments \cite{PhysRevB.64.214503,1367-2630-11-5-055055}. Out of the possible currents only two are not vanishing, the isospin and spin current densities. They are defined as follows:

\begin{align*}
    I^{(isospin)}(\vec k) =& -\frac{i}{\pi}\sum_\omega\left[\tr G_R(\omega,\vec k)\tau^1\right]
    =\cr
   =&-\frac{i}{\pi}\sum_\omega\sum_{I,J,\alpha}\tau^1_{IJ}\left[G_R(\omega,\vec k))\right]_{I\alpha,J\alpha}
\end{align*}
\begin{align*}
    I^{(spin)}(\vec k) =& -\frac{i}{\pi}\sum_\omega\left[\tr G_R(\omega,\vec k)i\Gamma^2\right]
    =\cr
  =&-\frac{i}{\pi}\sum_\omega\sum_{I,\alpha,\beta}(i\s^2)_{\alpha\beta}\left[G_R(\omega,\vec k))\right]_{I\alpha,I\beta}
\end{align*}
The local density of states $I(\vec k)=\sum_\omega \rho(\omega, \vec{k})$  is invariant under rotations in the symmetry plane. However the isospin and spin currents densities change as we rotate around the symmetry axis. In figure-\ref{fig:app-current} we plot the angle dependence of these currents for the $\ave{p_x}$ and $\ave{p_x+ip_y}$ superconductor. As expected from the symmetries of the background, the isospin current vanishes for the $\ave{p_x}$ type, but show a perfect sin behavior for the $\ave{p_x+ip_y}$ SC. The spin current is non-trivial for both cases.

Another interesting view, that give a better description on the location of the nodes is the spectral function in the momentum plane for fixed frequency. See figures \label{fig:app-momZ} and \label{fig:app-momW} for the two type of the superconductors. For the $\ave{p_x}$  we can see how the two nodes (peaks), appearing at the poles for low frequency. As the frequency increase the peaks move toward the center, and the spectral function turns into a Dirac cone. For the $\ave{p_x+ip_y}$ superconductor we see that the low frequency picture is a single node at $\theta = \pi$. As the frequency increase the node split into three peaks before developing into a Dirac cone.

\begin{figure}[h]
\begin{center}
\includegraphics[width=80mm]{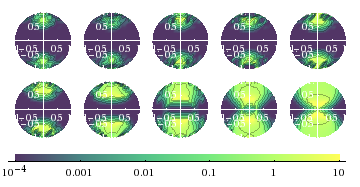}
\caption{\label{fig:app-momZ} The spectral function of a $\ave{p_x}$ superconductor as function of momenta for fixed frequency. Going from left to right the top row plots are for frequencies are $\omega = 0.025,0.05,0.075,0.01,0.125$, and the bottom row plots frequencies are $\omega = 0.15,0.225,0.3,0.375$ and $0.45$. We use a logarithmic color scale as described on the bottom of the plot.}
\includegraphics[width=80mm]{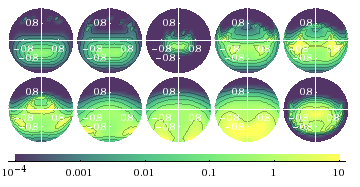}
\end{center}
\caption{\label{fig:app-momW} Spectral function of a $\ave{p_x+ip_y}$ SC as function of momenta for fixed frequency. See the caption of figure-\ref{fig:app-momZ} for details. }
\end{figure}

\bibliography{CSPbib}

\end{document}